
\magnification=\magstep1
\font\sixteenrm=cmr17
\def\a{\alpha}          \def\w{\omega}         \def\sy{\psi}
\def\g{\gamma}          \def\W{\Omega}         \def\Sy{\Psi}
\def\G{\Gamma}                   
                     
            \def\d{\delta}

                       \def\b{\beta}

\def\vo{\vartheta}                       
                     
\def\pd{\partial}

\def\syb{{\overline\psi}}
\def\Syb{{\overline\Psi}}
\def\N{\not\!\! N}
\def\^{\wedge}
\def\pd{\partial} 
\def\cd{\nabla}   

\nopagenumbers
\baselineskip=16 true pt plus1pt minus1pt
\line{\hfil gr-qc/9407004}
\bigskip
\centerline{\sixteenrm A Quadratic Spinor Lagrangian for General
Relativity}
\bigskip
\centerline{James M. Nester and Roh Suan Tung}
 \bigskip
\centerline{Department of Physics, National Central University}
\centerline{Chung-Li, Taiwan 32054, Republic of China}
\bigskip
\bigskip
\bigskip
\centerline{\underbar{Summary}}
\bigskip
\midinsert
\narrower\narrower
We present a new {\it finite} action for Einstein gravity in which
the Lagrangian is quadratic in the covariant derivative of a spinor
field. Via a new spinor-curvature identity, it is related to the
standard Einstein-Hilbert Lagrangian by a total differential term. The
corresponding Hamiltonian, like the one associated with the
Witten positive energy proof is fully four-covariant. It defines
quasi-local energy-momentum
and can be reduced to the one in our recent positive energy proof.
\endinsert

\bigskip
\bigskip
\vfil
\centerline{(Fourth Prize, 1994 Gravity
Research Foundation Essay.)}

\bigskip\bigskip
\vfil\eject
\pageno=1
\footline={\hfil\tenrm\folio\hfil}

The gravitational field responds to and exchanges energy-momentum with
its sources.  The matter and field sources are described by a proper
energy-momentum density tensor.  The energy-momentum density of
the gravitational field itself, however, has proved to be elusive.
To find a good mathematical expression for the gravitational energy-momentum
density has been a major impetus in the development of new techniques
for representing Einstein's gravity theory.

The prime principle of theoretical physics is to begin with
the action.  For Einstein's
General Relativity the usual starting point is
the Hilbert action in which the Lagrangian is the
scalar curvature.
But the scalar curvature contains second derivatives of the metric with an
associated asymptotically flat space fall off of $O(1/r^3)$ which
causes the action to diverge.
One can improve the situation while removing the second
derivatives of the metric by extracting a total differential.
However the remaining action is no longer covariant; consequently, the
energy momentum density constructed from it is a non-covariant object
--- a pseudotensor --- which can have any local value
 (including zero or {\it negative} values).
 After much effort along these lines many have given up on
the idea of constructing a reasonable gravitational energy-momentum density
by starting from the Lagrangian or in some other way.  Yet others find that
their physical intuition expects at least a quasi-local gravitational
energy-momentum density. Ideally it should be connected to the
action.

Here we present a finite action for Einstein gravity with a 4-dimensionally
covariant Lagrangian which leads to
a 4-covariant Hamiltonian which, in turn, permits not only a positive energy
proof but also a locally positive energy density as well as values
for quasi-local quantities.

We achieve this by introducing an auxiliary spinor field. The key is a newly
discovered spinor-curvature identity [1] which
permits the aforementioned extraction of a total differential to be done
in a covariant fashion.
In gravity research the efficacy
of spinor techniques (see e.g., [2,3,4]) has long been recognized
and
they have been used extensively for both utilitarian reasons as well as for
affording insights.  Nevertheless, in this widely explored
topic one can still discover new things.

In this work, we begin with a quadratic-spinor action which yields the Einstein
field equations. Then we obtain a covariant expression for the Hamiltonian.
This serves as our
energy-momentum density providing a positivity
proof and localization that has links with the Witten proof as well as others.


The new {\it quadratic spinor} action
is given by:
 $$    S[\Sy]=
 \int{\cal L} =
 \int 2 D\Syb \g_5 D\Sy
     \eqno(1) $$
where the gravitational variable is a spinor-valued 1-form
field $\Sy=\vo\sy$, which includes an orthonormal frame 1-form
$\vo:=\g_\a \vo^\a$ and
a normalized spinor field $\sy$ (i.e. $\syb\sy=1$).  The covariant
derivative
$D\Sy:=d\Sy+\w\Sy$
includes the matrix\footnote{${}^{\dag}$}{
The Dirac matrix conventions are $\g_{(\a}\g_{\b)}=g_{\a\b}$,
$\g_{\a\b}:=\g_{[\a}\g_{\b]}$, $\g_5:=\g^0\g^1\g^2\g^3$. The
metric signature is $(+---)$.}
 valued connection one-form
$\w:={{1\over4}}\g_{\a\b}\w^{\a\b}$.
(Clifford algebra valued forms [5] permit
 very succinct representations.)

The new {\it spinor-curvature identity} [1]
$$2D\Syb \g_5 D\Sy \equiv -\syb\sy\, R \ast\!1
+ d [(D\Syb)\g_5 \Sy+\Syb \g_5(D\Sy)] ,
\eqno(2) $$
reveals that
the quadratic-spinor Lagrangian
differs from the standard Einstein-Hilbert Lagrangian
only by a total differential term.  Hence they yield the same field
equations. However, the new quadratic spinor Lagrangian is asymptotically
$O(1/r^4)$ which guarantees finite action, an advantage over the
Einstein-Hilbert $O(1/r^3)$.

{}From the new Lagrangian, by variation with respect to $\Syb$,
we obtain Dimakis and M\"uller-Hoissen's [5] ``Clifform''
transcription of the (vacuum) Einstein field equations:
 $$
 {\d{\cal L}\over\d\Syb} = -2\g_5 D^2\Sy
  = -2 \g_5 \W\Sy
  = -\textstyle{1\over2}\W^{\a\b}\^\vo^\mu\g_5\g_{\a\b}\g_\mu\sy
  =  G_{\a\b} \ast\!\vo^{\a}\g^\b \sy=0,
\eqno(3)$$
where
$\W:=d\w+\w\^\w={{1\over4}}\g_{\a\b}\W^{\a\b}$
 is the matrix valued curvature 2-form.


The Hamiltonian can be constructed [6,7] from
 the action by choosing a
timelike evolution vector
field $N$ such  that $i_N dt = 1$ and splitting the action:
$S=\int {\cal L} =  \int dt \int i_N{\cal L}$.
This procedure yields the
Noether translation generator along $N$, i.e., the
 4-covariant Hamiltonian 3-form:
$$
{\cal H}(N)
=2[ D(\syb\N)\g_5 D(\vo\sy)+ D(\syb\vo)\g_5 D(\N\sy)].\eqno(4)$$
A notable feature is that
the Hamiltonian (4) is already $O(1/r^4)$.  Consequently, its integral will
be finite.  Moreover its variation will
have an $O(1/r^3)$ boundary term which will vanish asymptotically ---  there
is no need for a further adjustment by an additional
boundary term [8]. Indeed, although it is not readily apparent, the
Hamiltonian expression (4) could be obtained
from the usual (linear in the Einstein tensor) Hamiltonian by adding a
certain total differential
(although important for the value of energy-momentum such a total
differential does not effect the equations of motion) as the following
identity reveals:
$$ {\cal H}(N)\equiv-2\syb\sy N^\mu G_{\mu\nu}*\!\vo ^{\nu}+
2d[ \syb\N\g_5 D(\vo\sy)+ D(\syb\vo)\g_5 \N\sy].\eqno(5)
$$
This identity also shows that the derivatives of $\sy$ itself are not
so important --- up to an exact differential ${\cal H}(N)$ is algebraic
in $\sy$ ---
rather that these factors arrange for the correct quadratic connection terms.

The Hamiltonian 3-form ${\cal H}(N)$ is similar to
the one [6]
 $$\eqalignno{ {\cal H}_w &=2[ D\syb\g_5 D(\vo\sy) +
D(\syb\vo)\g_5 D\sy] \cr &\equiv -2N^\mu G_{\mu\nu}*\!\vo ^{\nu}+
2d[ \syb\vo\g_5 D\sy- D\syb\g_5\vo\sy], &(6) \cr}$$
associated with the Witten positive energy proof [3].  The principal
procedural difference is that the latter was obtained from the usual
Hamiltonian by reparameterization and discarding a boundary
term; unlike ${\cal H}(N)$ (4), it cannot be obtained from a 4-covariant
Lagrangian.  The reason for this lies in the principal technical
difference:  in eq (4) $N^{\mu}$ and $\sy$ are independent, whereas
in eq (6) the time evolution vector field,
$N^{\mu}=\syb\g^{\mu}\sy$, is tied to the spinor field.


The amount of energy-momentum within a region can be determined from
the value of the Hamiltonian.
With Einstein's
field equations satisfied, the Hamiltonian 3-form (5)
reduces to an exact differential which becomes an integral over the
2-surface bounding the region:
$$
E=H(N)_{\hbox{sol}}=\int\!{\cal H}(N)_{\hbox{sol}}
= \oint\! 2 N^\a
(\syb\g_5\g_{\a\b}\vo^\b\^D\sy- D\syb\^\g_5\vo^\b \g_{\b\a}\sy).
 \eqno(7)$$
This expression is comparable to the corresponding
Witten energy expression:
$$ E= -2\oint\!
(\syb\g_5\vo\^D\sy+ D\syb\^\g_5\vo\sy). \eqno(8)
$$

A variety of
expressions of this type have been used in quasi-local energy
investigations [9,10].  For any boundary values our new
4-covariant version (7) likewise
yields a quasi-local energy.  We have, as yet, no new ideas
 concerning the important
question of how to
select the best values on the boundary.  In addition to being 4-covariant the
new
expression has the merit of being connected with a Lagrangian.


Our new quadratic-spinor formulation permits a simple locally
positive
expression for the Hamiltonian density.
In eq (4) we
choose $\sy$ to conformally satisfy the Witten equation,
$    \g^a D_a (f\sy)=0 $, the shift vector to vanish and the
lapse $N=f^2$. A further algebraic
restriction on $\sy$ then reduces the Hamiltonian density to
$${\cal H}
=f^2 [4g^{ab} \cd_a\varphi^{\dag}\cd_b\varphi+K^{ab}K_{ab}-K^2
], \eqno(9)$$
which is the same simple locally positive (on maximal surfaces) expression
found in our recent positive energy proof.
It
has links to both the
{\it special orthonormal frame} (SOF) and the Witten proofs and
 their associated energy
localizations [11].


In summary we have
essentially presented a covariant version of the procedure of
removing a
total derivative from the Einstein-Hilbert action:
 $R\sim\pd\G+\G\G$.
We have used a spinor field to achieve this covariantization.
The quadratic spinor action considered here has an
$O(1/r^4)$ fall off which makes the action converge.
{}From it we obtained
a fully 4-covariant Hamiltonian which has an associated
well-defined boundary term that gives both total conserved and
quasi-local quantities for
gravitating systems.
The variables can be selected so that the Hamiltonian
 reduces to the one used in our recent positive energy
proof which has
links to both
the SOF and Witten approaches.

Perhaps it is worth remarking that
we have not given any physical interpretation to our spinor field.
The spinor field in our quadratic
spinor action is merely used here to provide a way of
 {\it covariantization} of the
$O(1/r^4)$ Lagrangian and Hamiltonian.

Of course there are
variations on our theme.  One can easily generalize to
the Einstein-Cartan theory by allowing non-vanishing torsion.
It is also possible to consider instead the (anti) self-dual connection
and curvature.  This would connect with work on New Variables.
Moreover, our new form of the action for Einstein's theory may yield
other benefits in addition to those considered here concerning
gravitational energy and its localization.
\bigskip
\beginsection ACKNOWLEDGEMENTS

We wish to thank Dr.~V.~V. Zhytnikov for helpful discussions and
the National Science Council of the R.O.C. for its support under
contract No. NSC 83-0208-M-008-014.
\beginsection REFERENCES

\frenchspacing

\item{1.} J. M. Nester , R. S. Tung and  V. V. Zhytnikov,
         {\it Class. Quantum Grav. \bf 11}, 983 (1994).

\item{2.} A. Ashtekar, {\it Phys. Rev. D \bf 36}, 1587 (1987).

\item{3.} E. Witten,
         {\it Comm. Math. Phys. } {\bf 80}, 381 (1981).

\item{4.} R. Penrose and W. Rindler,
 {\it Spinors and Space-Time},
 (Cambridge Univ. Press, 1986).

\item{5.} A. Dimakis and F. M\"uller-Hoissen,
      {\it Class. Quantum Grav.} {\bf 8}, 2093 (1991).

\item{6.}  J. M.  Nester,
       {\it Phys. Lett. A \bf 83}, 241 (1981);
         in {\it Asymptotic Behavior of Mass and Space-time Geometry},
         ed F Flaherty
         (Springer, Berlin, 1984) p.155 .

\item{7.} J. M. Nester, {\it Mod. Phys. Lett. A} {\bf 6} 2655 (1991);
                in {\it Directions in General Relativity,}
                Vol. I, ed. B.L. Hu, M.P. Ryan and C.V. Vishveshwara,
                (Cambridge Univ. Press, 1993).

\item{8.}   T. Regge and C. Teitelboim,
            {\it Ann. Phys.} (NY) {\bf 88}, 286 (1974).

\item{9.}   G. Bergqvist, {\it Class. Quantum Grav.} {\bf 9}, 1753
            (1992).

\item{10.}   A. J. Dougan and L. J. Mason, {\it Phys. Rev. Lett.} {\bf
             67}, 2119 (1991).

\item{11.} J. M. Nester and R. S. Tung,
           {\it Phys. Rev. D \bf 49}, 3958 (1994).

\vfill\eject
\bye